\documentclass[Physsubmission, Phys]{SciPost}

\hypersetup{
    colorlinks,
    linkcolor={red!50!black},
    citecolor={blue!50!black},
    urlcolor={blue!80!black}
}

\usepackage[bitstream-charter]{mathdesign}
\DeclareSymbolFont{usualmathcal}{OMS}{cmsy}{m}{n}
\DeclareSymbolFontAlphabet{\mathcal}{usualmathcal}

\usepackage{graphicx}
\usepackage{mathtools}
\usepackage{enumerate}

\graphicspath{{../J_psi_ccbarg/old/}}

\newcommand{\bq}{\mbox{\boldmath $q$}}
\newcommand{\br}{\mbox{\boldmath $r$}}

\newcommand{\brho}{\mbox{\boldmath $\rho$}}

\newcommand{\bb}{\mbox{\boldmath $b$}}

\newcommand{\ket}[1]{| {#1} \rangle}
\newcommand{\bra}[1]{\langle {#1} |}

\newcommand{\half}{{1\over 2}}

\def\lsim{\mathrel{\rlap{\lower4pt\hbox{\hskip1pt$\sim$}}
		\raise1pt\hbox{$<$}}}         
\def\gsim{\mathrel{\rlap{\lower4pt\hbox{\hskip1pt$\sim$}}
		\raise1pt\hbox{$>$}}}         

\begin{document}

\begin{center}{\Large \textbf{
Coherent photoproduction of $J/\psi$ in nucleus-nucleus collisions \\in the color dipole approach- an update\\
}}\end{center}

\begin{center}
Agnieszka {\L}uszczak\textsuperscript{1$\star$} and
Wolfgang Sch{\"a}fer\textsuperscript{2} 
\end{center}

\begin{center}
{\bf 1} Cracow University of Technology, PL-31155 Cracow, Poland\\
{\bf 2} Institute of Nuclear Physics Polish Academy of Sciences, PL-31342 Cracow, Poland
\\
* agnieszka.luszczak@pk.edu.pl
\end{center}

\begin{center}
\today
\end{center}


\definecolor{palegray}{gray}{0.95}
\begin{center}
\colorbox{palegray}{
  \begin{tabular}{rr}
  \begin{minipage}{0.1\textwidth}
    \includegraphics[width=22mm]{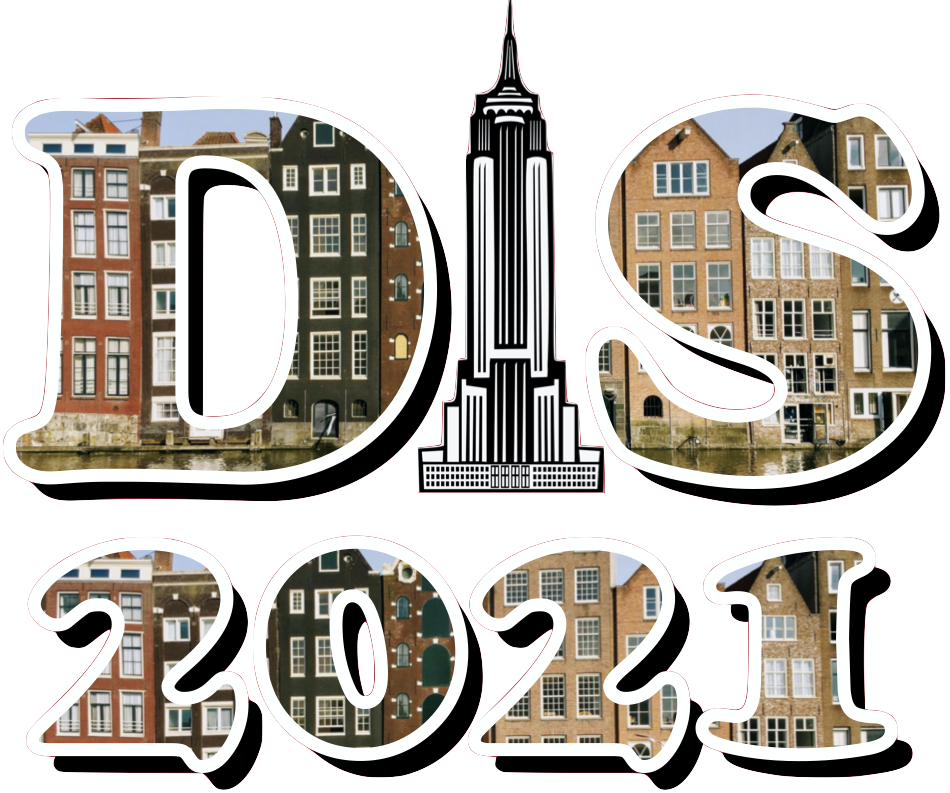}
  \end{minipage}
  &
  \begin{minipage}{0.75\textwidth}
    \begin{center}
    {\it Proceedings for the XXVIII International Workshop\\ on Deep-Inelastic Scattering and
Related Subjects,}\\
    {\it Stony Brook University, New York, USA, 12-16 April 2021} \\
    \doi{10.21468/SciPostPhysProc.?}\\
    \end{center}
  \end{minipage}
\end{tabular}
}
\end{center}

\section*{Abstract}
{\bf
We discuss the role of $c \bar c g$-Fock states in the diffractive photoproduction
of $J/\psi$-mesons. We build on our earlier description of the process 
in the color-dipole approach, where we took into account the rescattering of
$c \bar c$ pairs using a Glauber-Gribov form of the dipole-nucleus amplitude.
We compare the results of our calculations to recent data on the
photoproduction of $J/\psi$ by the ALICE and LHCb collaborations.
}



\section{Introduction}
\label{sec:intro}

Recent measurements
\cite{Abelev:2012ba,Abbas:2013oua,Khachatryan:2016qhq,Kryshen:2017jfz,LHCb:2018ofh}
(see also the review \cite{Contreras:2015dqa}) 
of exclusive 
production of $J/\psi$ mesons in ultraperipheral heavy-ion 
collisions at the LHC have given us new access to the interaction of small color dipoles with cold nuclear matter. 

Indeed, in the limit of large photon energy $\omega$ in the rest frame of the nucleus,  the coherence length $l_c = 2 \omega /M_V^2$ for the vector meson of mass $M_V$ becomes much larger than the size of the nucleus $l_c \gg R_A$ \cite{Kopeliovich:1991pu,Nikolaev:1992si}. 
The photoproduction of the $J/\psi$ meson can then be described as a splitting of the photon 
into a $c \bar c$ pair far upstream the target, and an interaction of a color dipole of size
$\br$ formed by quark and antiquark. The scattered
$c \bar c$ pair then evolves into the final state
vector meson.
The dominantly imaginary forward amplitude of interest then takes the form
\begin{eqnarray}
\mathcal{A}(\gamma A \rightarrow V A ;W,\bq=0)  &=& 
2i \int d^2\bb \,  \bra{V} \Gamma_A(x,\bb,\br) \ket{\gamma} \nonumber \\
&=& 2i \int_0^1 dz \int d^2\br \, \Psi^*_V(z,\br) \Psi_\gamma(z,\br)  \, \int d^2 \bb \, \Gamma_A(x,\bb,\br).
\label{eq:amplitude}
\end{eqnarray}
Here $W$ is the $\gamma A$ per-nucleon cm-energy, and $x = M_V^2/W^2$. By $z$ we denote the lightcone momentum fraction of the photon momentum carried by the quark. The $c \bar c$-Fock state light-front wave functions of photon and vector meson are denoted by $\Psi_\gamma$ and $\Psi_V$ respectively, and we suppressed a sum over the quark/antiquark helicities, which are conserved in the interaction with the target.

\begin{figure}[!h]
	\begin{center}
	a) \includegraphics[width=.45\textwidth]{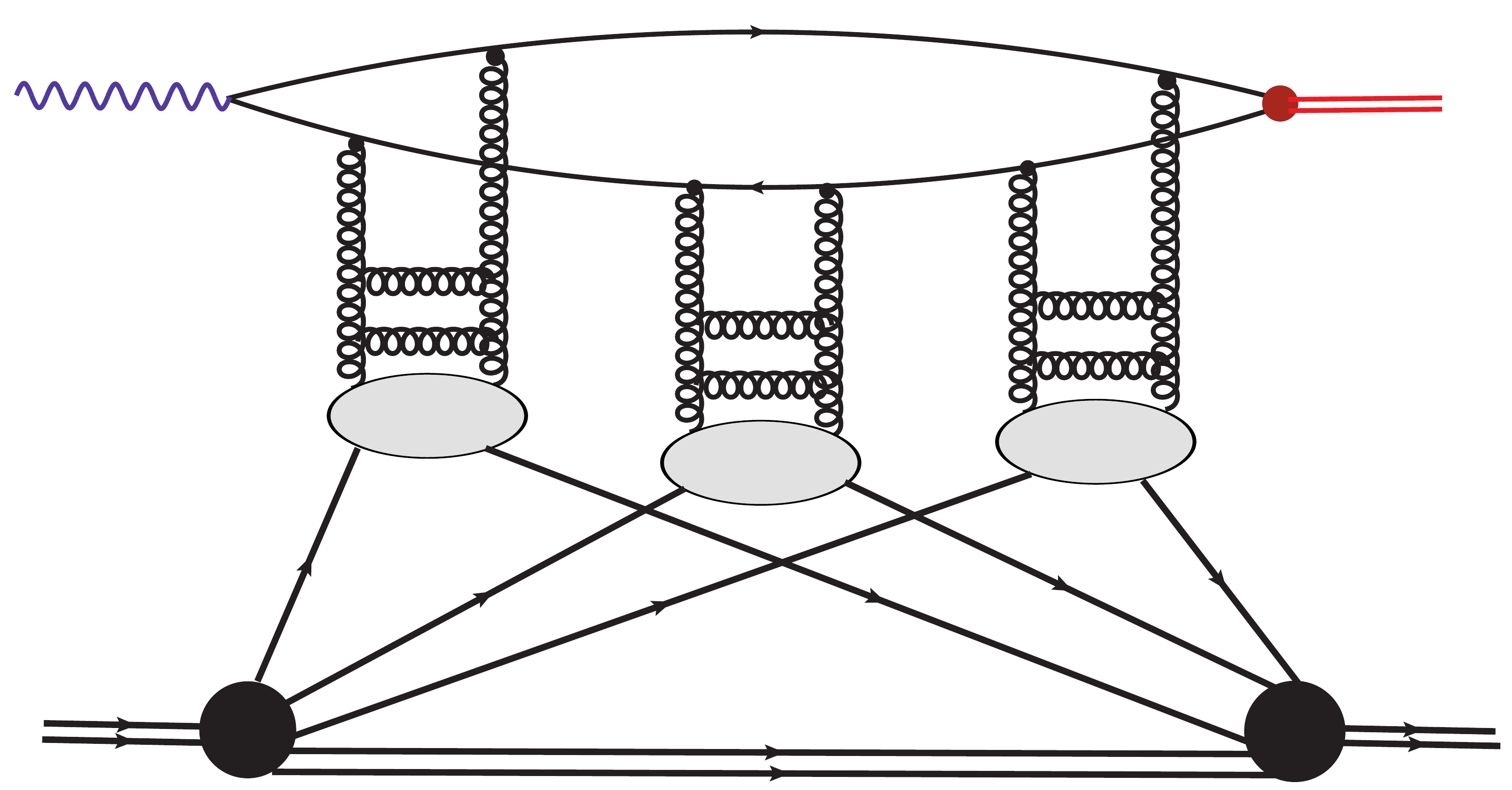}
	b)	\includegraphics[width=.45\textwidth]{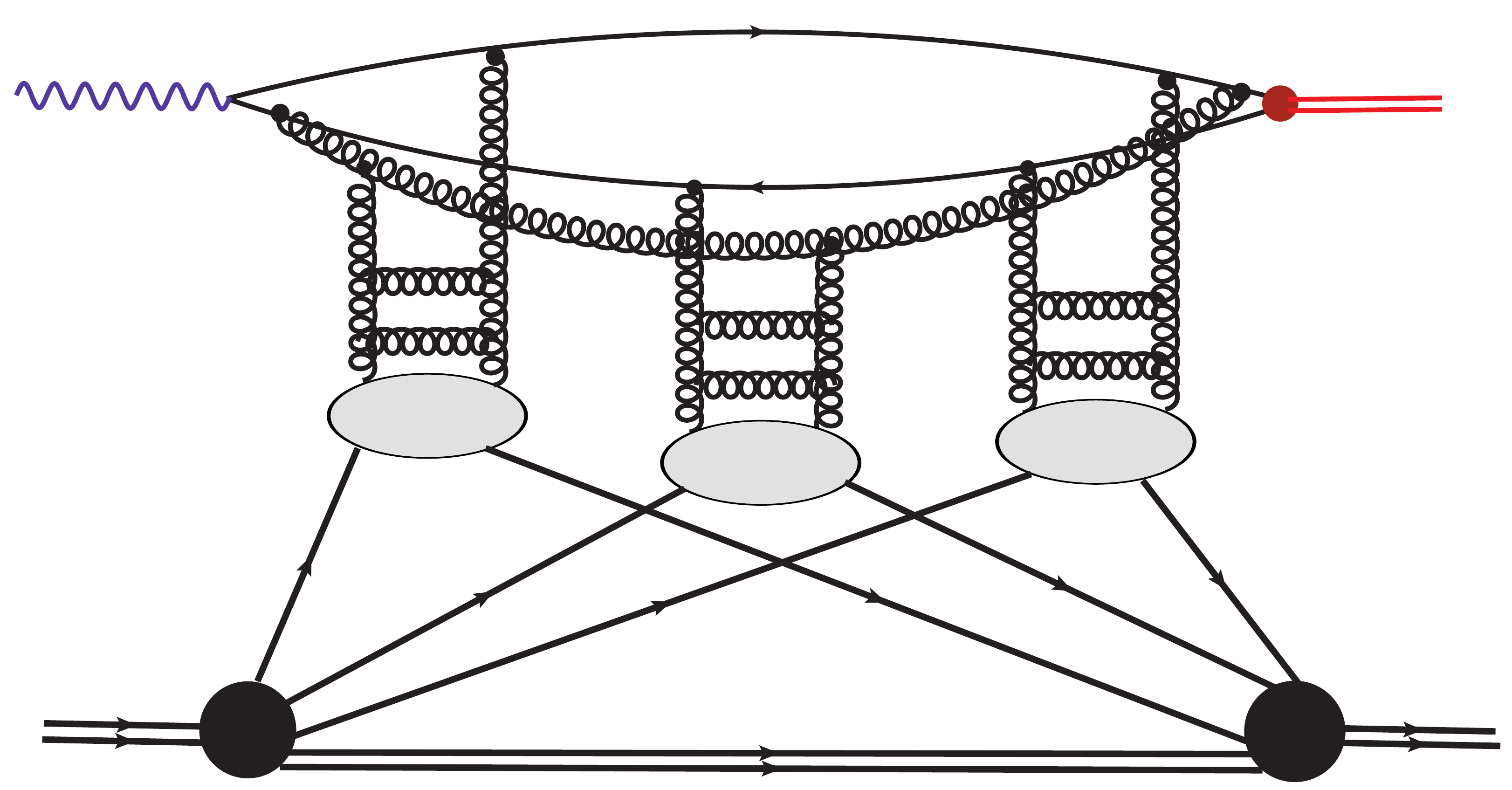}
		\caption{Coherent photoproduction of a vector meson in which the nucleus stays in its ground state.  
		}
		\label{fig:diagrams}
	\end{center}
\end{figure}

Here we continue our investigations \cite{Luszczak:2017dwf,Luszczak:2019vdc}, with a nuclear dipole cross section which is based on its free-nucleon counterpart obtained through fits to HERA data \cite{Luszczak:2013rxa,Luszczak:2016bxd}.\\
In Refs.\cite{Luszczak:2017dwf,Luszczak:2019vdc}, we used the dipole-nucleus amplitudes obtained from applying the rules of an extended Glauber-theory to color dipoles as a set of eigenstates of the scattering \cite{Nikolaev:1990ja}. In particular, the dipole-nucleus amplitude in
impact parameter space is obtained as 
\cite{Nikolaev:1990ja,Mueller:1989st}:
\begin{eqnarray}
\Gamma_A(x,\bb,\br) = 1 -  S_A(x,\bb,\br)\, , \, \, {\rm{with} }\, \, S_A(x,\bb,\br) = \exp\Big[-\half \sigma(x,\br) T_A(\bb)\Big] \, .
\label{eq:Glauber}
\end{eqnarray}
Above, $T_A(\bb) = \int_{-\infty}^\infty \, dz n_A(\sqrt{\bb^2 +z^2})$ is the optical thickness of the nucleus of mass number $A$ at impact parameter $\bb$, with the nuclear matter density $n_A(R)$ being normalized as 
$\int d^3{\vec R} \, n_A(R) = A$. The formula Eq.\ref{eq:Glauber} corresponds to a summation of diagrams of the type shown in Fig.~\ref{fig:diagrams}a. It takes into account the multiple scattering of the $c \bar c$-dipole on the constituent protons and neutrons of the nucleus. 

In the midrapidity region the maximum of the $\gamma A$-cm energy accessible in the collision is obtained. Roughly we have there $W \sim 100 \, \rm{GeV}$.
With increasing energy, the coherency condition $l_c \gg R_A$ will be satisfied not only by the $c \bar c$-state, but also by higher $c \bar c g$ states shown in in Fig.~\ref{fig:diagrams}b.
In the language of Glauber--Gribov theory, these correspond to 
inealastic shadowing corrections induced by high--mass diffractive states.

In this work we wish to address the possible role of these high mass states, restricting ourselves to the $c \bar c g$ component.

\section{Contribution of the $c\bar c g$~Fock state}

In this section we briefly review how higher Fock-states are accounted for in the color-dipole formalism. For the problem at hand, the Fock-state expansion of the photon 
reads, schematically
\begin{eqnarray}
\ket{\gamma} = \sqrt{Z_g} \Psi_{c \bar c} \ket{c \bar c} + \Psi_{c \bar c g} \ket{c \bar c g} + \dots \, .
\end{eqnarray}
Here $\Psi_{c \bar c}, \Psi_{c \bar c g}$ are the light-front wavefunctions (WFs) of the two- and three-body Fock states respectively. Virtual corrections induce the renormalization of the $c \bar c$ state by the (formally divergent) factor $\sqrt{Z_g}$.
For gluons which carry a small light-cone momentum fraction $z_g \ll 1$, the three-body WF takes a factorized form, $\Psi_{c \bar c g} = \Psi_{c \bar c} \, ( \Psi_{cg} - \Psi_{\bar c g} )$.

To evaluate the effect of the $c \bar c g$-state on the nuclear amplitude, we still need the cross section of the three-body system with the nucleon. As for the $c \bar c$-dipole, the impact parameters and helicities of partons in the Fock-state are conserved. Let us denote the $c$-$g$ and $\bar c$-$g$ transverse distances by $\brho_1$ and $\brho_2$, respectively, and the $c \bar c$ separation by $\br = \brho_1 - \brho_2$. 
Then, following  Refs.\cite{Nikolaev:1993ke,Nikolaev:1994kk,Nikolaev:1993th}, the dipole cross section for the three-body system is
\begin{eqnarray}
	\sigma_{q \bar q g}(x,\brho_1,\brho_2,\br) = {C_A \over 2 C_F} \Big( \sigma(x,\brho_1) + \sigma(x,\brho_2) - \sigma(x,\br) \Big) + \sigma(x,\br) \, ,
	\label{eq:sigma_qqbarg}
\end{eqnarray}
where $C_A=N_c$ and $C_F = (N_c^2 -1)/(2 N_c)$ are the standard Casimirs for the color-$SU(N_c)$ adjoint and fundamental represenations.
In the limit of small $c \bar c$ separation, $\br \to 0, \brho_1 \to \brho_2 \equiv \brho$, the $q \bar q g$ cross section approaches $\sigma_{q \bar q g} \to {C_A \over C_F} \sigma(x,\brho)$, which is precisely the dipole cross section for the dipole formed out of two adjoint color charges (gluons).
The nuclear $S$-matrix for the $c \bar c g$-state would now be obtained from applying the Glauber-form to the cross section Eq.(\ref{eq:sigma_qqbarg}). In a large-$N_c$ approximation, the three-body $S$-matrix factorizes as
\begin{eqnarray}
S_{q\bar q g, A}(x,\brho_1,\brho_2,\bb) = S_A(x,\brho_1, \bb + {\brho_2 \over 2}) S_A(x,\brho_2, \bb + {\brho_1 \over 2}) 
\end{eqnarray}
Taking due care of the virtual correction to the two-body Fock state, after integrating over degrees of freedom of the gluon, we obtain the full dipole-nucleus amplitude as:
\begin{eqnarray}
	\Gamma_A (x,\br,\bb) = \Gamma_A(x_A,\br,\bb) + \log\Big({x_A \over x}\Big) \Delta\Gamma_A(x_A,\br,\bb)  \, ,
	\label{eq:Gamma_full}
\end{eqnarray}
with the correction to the dipole-amplitude from the $c \bar c g$ state:
\begin{eqnarray}
	\Delta \Gamma_A (x_A,\br,\bb) &=& \int d^2\brho_1 |\psi(\brho_1) - \psi(\brho_2)|^2 \Big\{ \Gamma_A(x_A,\brho_1,\bb+{\brho_2 \over 2}) 
	+ \Gamma_A(x_A,\brho_2,\bb+{\brho_1 \over 2}) 
	\nonumber \\ 
	&&- \Gamma_A(x_A,\br,\bb)  - \Gamma_A(x_A,\brho_1,\bb+{\brho_2 \over 2}) \Gamma_A(x_A,\brho_2,\bb+{\brho_1 \over 2}) \Big\}  \, ,
	\label{eq:Delta_Gamma}
	\end{eqnarray}
Here the logarithm in Eq.(\ref{eq:Gamma_full}) comes from the integration over the longitudinal phase-space of the gluon, 
where the WF of the gluon with $z_g \ll 1$ leads to the $dz_g/z_g$ integration.
There remains a dependence on transverse separation of the gluon from quark/antiquark, encoded in the radial WF:
\begin{eqnarray}
\psi(\brho) = {\sqrt{C_F \alpha_s} \over \pi} 
{\brho \over \rho R_c} K_1\Big({\rho \over R_c}\Big) \, 
\end{eqnarray}
In Eq.(\ref{eq:Delta_Gamma}) the integration extend over all dipole sizes, including the infrared domain of large dipoles, where perturbation theory does not apply. Here, we follow \cite{Nikolaev:1994vf,Nikolaev:2006mh}. by introducing the minimal  regularization of pQCD in terms of the finite propagation radius $R_c \sim 0.2 \div 0.3 \, \mathrm{fm}$ accompanied by a corresponding freezing of $\alpha_s$ in the infrared.

In order to quantify the nuclear suppression of coherent diffractive production, we write the nuclear cross section as
\begin{eqnarray}
\sigma(\gamma A \to J/\psi A; W) = R_{\rm coh}(x) \, \sigma(\gamma p \to J/\psi p; W)  \, .
\label{eq:total_cross_sec}
\end{eqnarray}
Here $R_{\rm coh}$ is evaluated as:
\begin{eqnarray}
R_{\rm coh}(x) = {\displaystyle \int d^2\bb \, \Big| \bra{J/\psi} \Gamma_A(x,\br, \bb) \ket{\gamma}\Big|^2 \over
\displaystyle \int d^2\bb \, \Big| \bra{J/\psi} \Gamma_{\rm IA}(x,\br, \bb) \ket{\gamma}\Big|^2 \, ,}
\end{eqnarray}
with the impulse approximation in the denominator at $x=x_A$ being defined as
\begin{eqnarray}
\Gamma_{\rm IA}(x_A,\br,\bb) = \half \sigma(x_A,\br) T_A(\bb) \, ,
\end{eqnarray}
which is then inserted into Eq.(\ref{eq:Delta_Gamma}) with the nonlinear term omitted for consistency.

The cross section $\sigma(\gamma p \to J/\psi p; W)$ appearing in Eq.(\ref{eq:total_cross_sec}) is taken from our previous work \cite{Luszczak:2019vdc}.

\section{Numerical results}
\label{sec:results}
In our numerical calculations we use the same light-front wave function as used in \cite{Luszczak:2019vdc}, and the dipole cross section obtained in \cite{Luszczak:2016bxd}. We refer the reader to these references to details which must not be repeated here.

\begin{figure}[!h]
	\begin{center}
	\includegraphics[width=.45\textwidth]{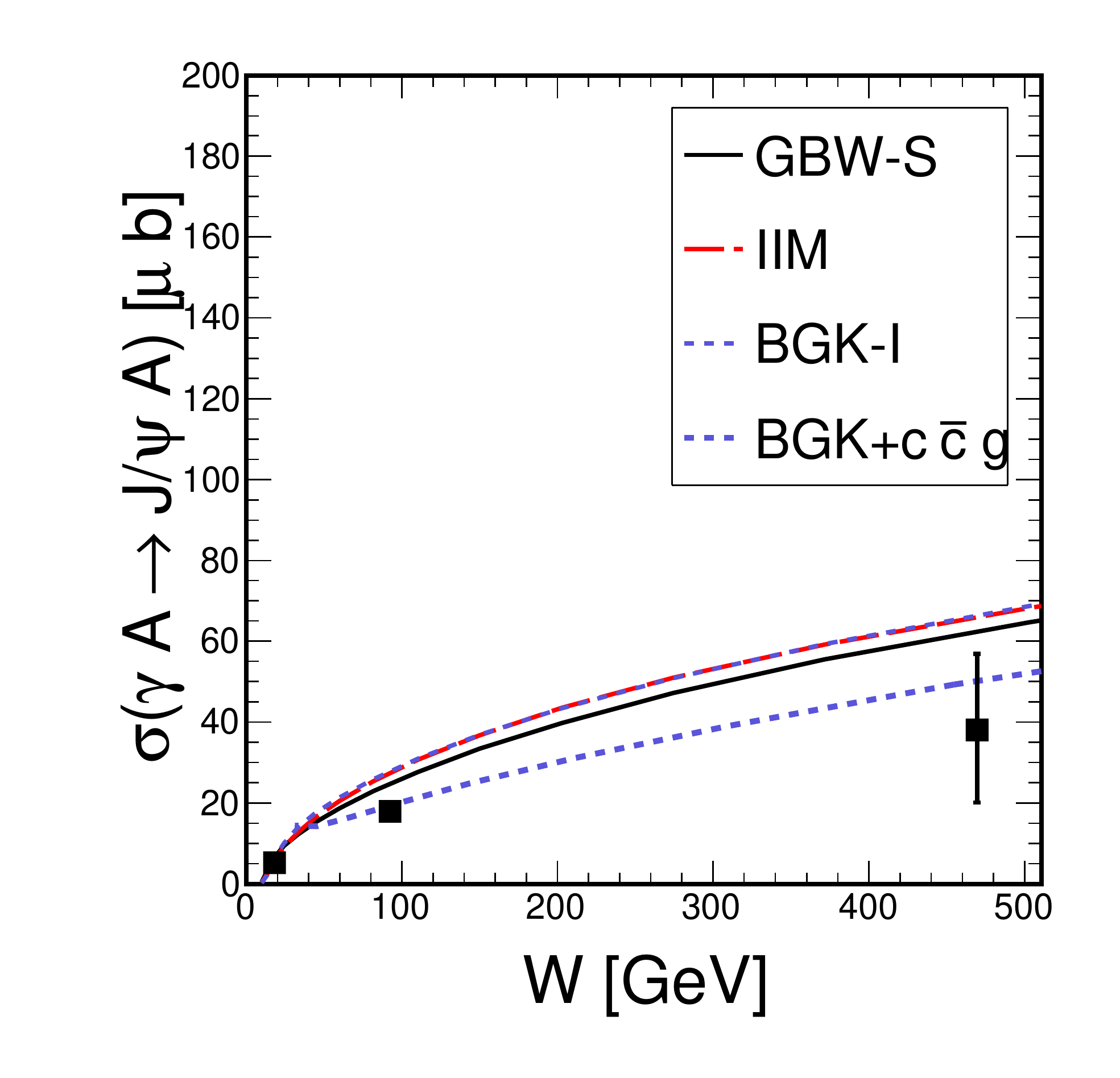}
		\caption{The total cross section for the diffractive photoproduction of $J/\psi$ on the lead nucleus. The data points are taken from Ref.\cite{Contreras:2016pkc}.  
		}
		\label{fig:sigma_A}
	\end{center}
\end{figure}

In Fig.\ref{fig:sigma_A} we show our results for the total diffractive photoproduction cross section of $J/\psi$ on lead as a function of $\gamma A$ per-nucleon cm-energy.
The data points were extracted by Contreras \cite{Contreras:2016pkc} from
data obtained in ultraperipheral heavy-ion collisions.
We observe that the calculations including the effect of the $c \bar cg$ state show an additional suppression of the nuclear cross section, as required
by experimental data.

We now wish to compare directly to the rapidity-dependent cross sections for ultraperipheral lead-lead collision.
To this end we use the standard Weizs\"acker-Williams approximation
\begin{eqnarray}
{d \sigma (AA \to AA J/\psi;\sqrt{s_{NN}}) \over dy} = 
n(\omega_+) \sigma(\gamma A \to J/\psi A; W_+) + n(\omega_-) \sigma(\gamma A \to J/\psi A; W_-) \, . \nonumber \\
\end{eqnarray}
We use the standard form of the Weizs\"acker-Williams flux (see e.g. the reviews \cite{Bertulani:2005ru,Baur:2001jj}) for the ion
moving with boost $\gamma$:
\begin{eqnarray}
n(\omega) = {2 Z^2 \alpha_{\rm em} \over \pi} \Big[ \xi K_0(\xi) K_1(\xi) - {\xi^2 \over 2} (K_1^2(\xi) -K_0^2(\xi))  \Big].
\end{eqnarray}
Here $\omega$ is the photon energy, and $\xi = 2R_A \omega/\gamma$. This flux was obtained by imposing the constraint
on the impact parameter of the collision $b > 2R_A$, where we use $R_A= 7 \,\rm{fm}$.
Here $\omega$ is the photon energy, and $\xi = 2R_A \omega/\gamma$. This flux was obtained by imposing the constraint
on the impact parameter of the collision $b > 2R_A$, where we use $R_A= 7 \,\rm{fm}$.
This means that configurations where nuclei touch each other are excluded, as otherwise inelastic processes would destroy
the rapidity gaps in the event.
The photon energies corresponding to the two contributions are
$\omega_\pm = m_V \exp[\pm y]/2$, the corresponding cms-energies for the $\gamma A \to J/\psi A$ subprocesses
are $W_\pm = 2 \sqrt{s_{NN}} \omega_\pm$.

\begin{figure}[!h]
	\begin{center}
	a) \includegraphics[width=.45\textwidth]{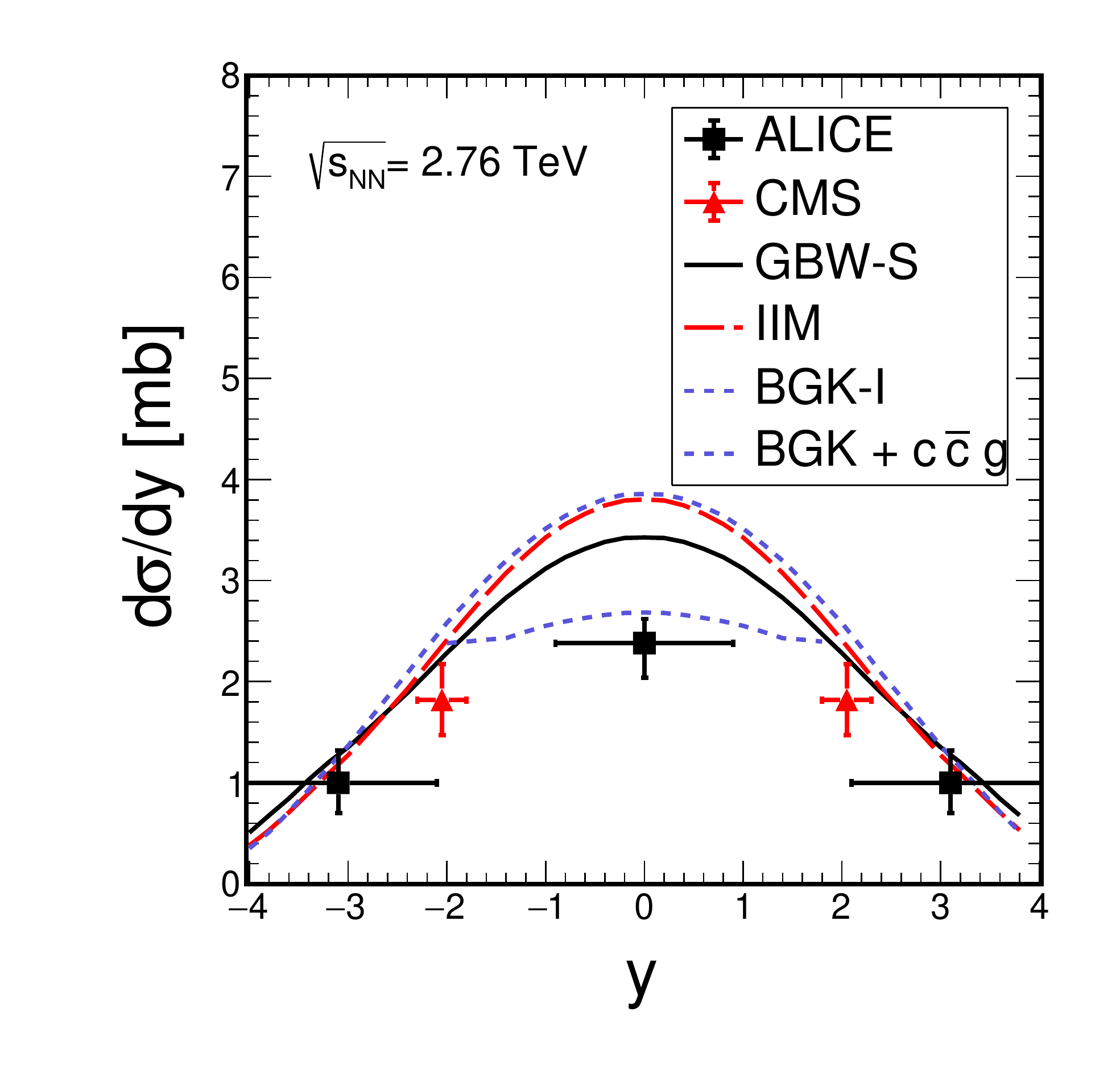}
	b)	\includegraphics[width=.45\textwidth]{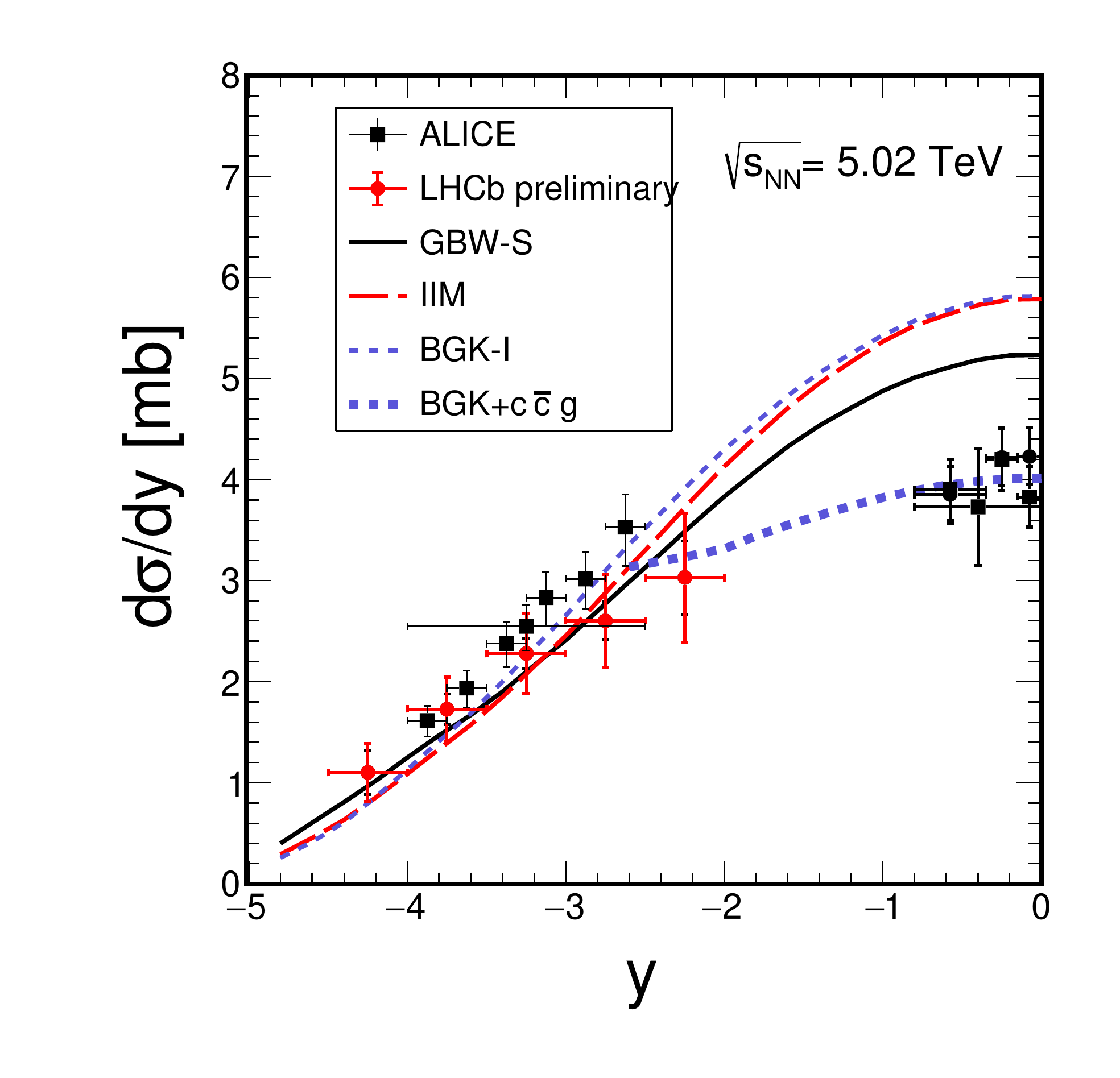}
		\caption{Rapidity dependent cross section fro the coherent photoproduction of $J/\psi$ in lead-lead collisions for two different energies. The thick dashed curve contains the $c \bar c g$-state with $R_c = 0.215$ fm.   
		}
		\label{fig:dsig_dy}
	\end{center}
\end{figure}

\newpage
In Fig.\ref{fig:dsig_dy}a) we compare
to data of ALICE and CMS at $\sqrt{s_{NN}} = 2.76 \, \rm{TeV}$, 
while Fig.\ref{fig:dsig_dy}b) we show the comparison with data of LHCb and ALICE at $\sqrt{s_{NN}} = 5.02 \, \rm{TeV}$, .

\section{Conclusions}

In this analysis we have demonstrated, that the inclusion of inelastic shadowing due to high-mass diffractive states leads to an additional suppression
of the coherent $J/\psi$ photoproduction on lead.
We modeled the diffractively excited high-mass system by the $c \bar c g$ Fock state of the photon. We observe that the inclusion of $c \bar c g$-states improves agreement of the dipole approach with the midrapidity data of the ALICE collaboration. Admittedly, there is a sizeable dependence on the gluon correlation radius $R_c$, which means that a calculation in a purely perturbative approach is not viable. Here we show calculations for with $R_c = 0.215$ fm.

We believe that our modeling of the essentially nonperturbative physics is well motivated by a phenomenological success of earlier works in the color dipole approach , e.g. \cite{Nikolaev:1994vf,Nikolaev:2006mh}.
A restriction to the $c \bar c g$-system is backed up by the fact, that diffractive structure functions of the proton measured at HERA are well described by the inclusion of $q \bar q$ and $q \bar q g$-states \cite{Golec-Biernat:2008mko}.



 \section*{Acknowledgements}
\paragraph{Funding information}
The  work  reported here was  partially  supported  by  the  Polish  National  Science  Center (NCN) grant  UMO-2018/31/B/ST2/03537.





\end{document}